\documentclass{ws-procs9x6}

\def\CH{{\mbox{\scriptsize{CH}}}}

\def\mi{{\mbox{\scriptsize{min}}}}
\def\ma{{\mbox{\scriptsize{max}}}}
\def\eV{{\mbox{eV}}}

\def\mbb{\langle m \rangle_{\beta\beta}}
\def\0nbb{0\nu\beta\beta}

\begin{document}

\title{
Neutrinoless Double Beta Decay Constraints%
\footnote{
\uppercase{I}nvited talk at \uppercase{NOON}2003,
\uppercase{F}eb.\ 10-14, 2003, \uppercase{K}anazawa, \uppercase{J}apan.
}
}

\author{
Hiroaki Sugiyama
}

\address{Department of Physics, Tokyo Metropolitan University,\\
Hachioji, Tokyo 192-0397, Japan\\
E-mail: hiroaki@phys.metro-u.ac.jp}


\maketitle

\abstracts{
 A brief overview is given of
theoretical analyses with neutrinoless double beta decay experiments.
 Theoretical bounds on the ``observable'', $\mbb$, are presented.
 By using experimental bounds on $\mbb$,
allowed regions are obtained on the \mbox{$m_l$-$\cos{2\theta_{12}}$} plane,
where $m_l$ stands for the lightest neutrino mass.
 It is shown that Majorana neutrinos can be excluded
by combining possible results of future neutrinoless double beta decay
and ${}^3H$ beta decay experiments.
 A possibility to constrain one of two Majorana phases is discussed also. 
}

\section{Introduction}

 Neutrino oscillation experiments so far supplied us
with much knowledge of the lepton flavor mixing.
 As we know more about that,
our desires of knowing better grows also about
what can not be determined by oscillation experiments:
 How heavy are neutrinos?
 Are they Majorana particles?
 The former question can be answered directly
by ${}^3H$ beta decay experiments.
 On the other hand,
neutrinoless double beta decay ($\0nbb$) experiments can give
the direct answer to the latter one.
 $\0nbb$ experiments seem to be interesting especially
because some of the future $\0nbb$ experiments are expected to probe
rather small energy scale $\sim10^{-2}\,\eV$,
and then will give stringent constraints on
neutrino masses and mixing parameters.

 The constraints by $\0nbb$ experiments
have been discussed by many authors.%
\footnote{For example, see references%
\cite{Petcov:1993rk,Minakata:1996vs,Adhikari:1998du,Bilenky:1999wz,Czakon:1999cd,Vissani:1999tu,Klapdor-Kleingrothaus:2000gr,Farzan:2001cj,Pascoli:2001by,Rodejohann:2002ng,Matsuda:2002iw,Barger:2002vy,Nunokawa:2002iv,Pascoli:2002qm,Minakata:2002ts}
and the references therein.}
 In this talk,
I will try to give a brief overview of those in a manner
as clear as possible.
 For generality and to emphasize future perspective,
we will not use any specific results
of $\0nbb$ experiments.
 Specific constraints can be obtained easily
by replacing $\mbb^\mi$ and $\mbb^\ma$,
which denote experimental bounds, in equations
with actual results of $\0nbb$ experiments.

\section{Theoretical constraints on $\mbb$\label{secttheo}}
 In the standard parametrization the MNS matrix for three neutrinos is
 \begin{eqnarray}
  U_{\mbox{\scriptsize{MNS}}}
   \equiv
   \left[
    \begin{array}{ccc}
     c_{12}c_{13} & s_{12}c_{13} & s_{13}e^{-i\delta}\\
     -s_{12}c_{23}-c_{12}s_{23}s_{13}e^{i\delta} &
      c_{12}c_{23}-s_{12}s_{23}s_{13}e^{i\delta} & s_{23}c_{13}\\
     s_{12}s_{23}-c_{12}c_{23}s_{13}e^{i\delta} &
      -c_{12}s_{23}-s_{12}c_{23}s_{13}e^{i\delta} & c_{23}c_{13}
    \end{array}
   \right].
   \label{MNSmatrix}
 \end{eqnarray}
 In most parts of this talk we assume that neutrinos are Majorana particles.
 Then, the mixing matrix includes two extra CP-violating phases
 (Majorana phases) as
 \begin{equation}
  U \equiv U_{\mbox{\scriptsize{MNS}}}
   \times \mbox{diag}(1, e^{i\beta}, e^{i\gamma}).
 \end{equation}
 The ``observable'' of $\0nbb$ experiments
\footnote{The actual observable is the half life
$T_{1/2}^{0\nu}
 = \left( G^{0\nu}\,|M^{0\nu}|^2\,\mbb^2 \right)^{-1}$
with the nuclear matrix elements $M^{0\nu}$
and the calculable phase space integral $G^{0\nu}$.} is
 \begin{eqnarray}
  \mbb
   \equiv
   \left|\, \sum^{3}_{i=1} m_i U^2_{ei}\, \right|
   =
   \left|\,
    m_1 c_{12}^2 c_{13}^2
    + m_2 s_{12}^2 c_{13}^2 e^{2 i \beta}
    + m_3 s_{13}^2 e^{2 i (\gamma - \delta)}\,
   \right| ,
   \label{beta1}
 \end{eqnarray}
where $U_{ei}$ represent the elements in the first low of U
and $m_i > 0$ are the neutrino mass eigenvalues.
 The experimental constraints on mixing angles are
$\sin^2{2\theta_{13}} \lesssim 0.1 \equiv \sin^2{2\theta_\CH}$
of the CHOOZ bound
and $0.2 < \cos{2\theta_{12}} < 0.5$ ($0.4 > s^2_{12} > 0.25$)
with the best fit $\cos{2\theta_{12}} = 0.37$ ($s_{12}^2 = 0.315$).
 For simplicity, $\Delta m^2_{ij} \equiv m^2_j - m^2_i$ are fixed
as $\Delta m^2_{12} = 7.3 \times 10^{-5}\,\eV$
and $|\Delta m^2_{23}| = 2.5 \times 10^{-3}\,\eV$.
 Hereafter, we call $m_1 < m_2 < m_3$ case the normal hierarchy
and $m_3 < m_1 < m_2$ case the inverted hierarchy
even if masses are almost degenerate.
 The lightest mass eigenvalue $m_l$,
which is $m_1$($m_3$) for the normal (inverted) hierarchy,
is used as the horizontal axis of all figures in this talk
to avoid tedious case studies for
whether the mass pattern is hierarchical or not.

\begin{figure}[t]
 \includegraphics[width=0.48\textwidth]{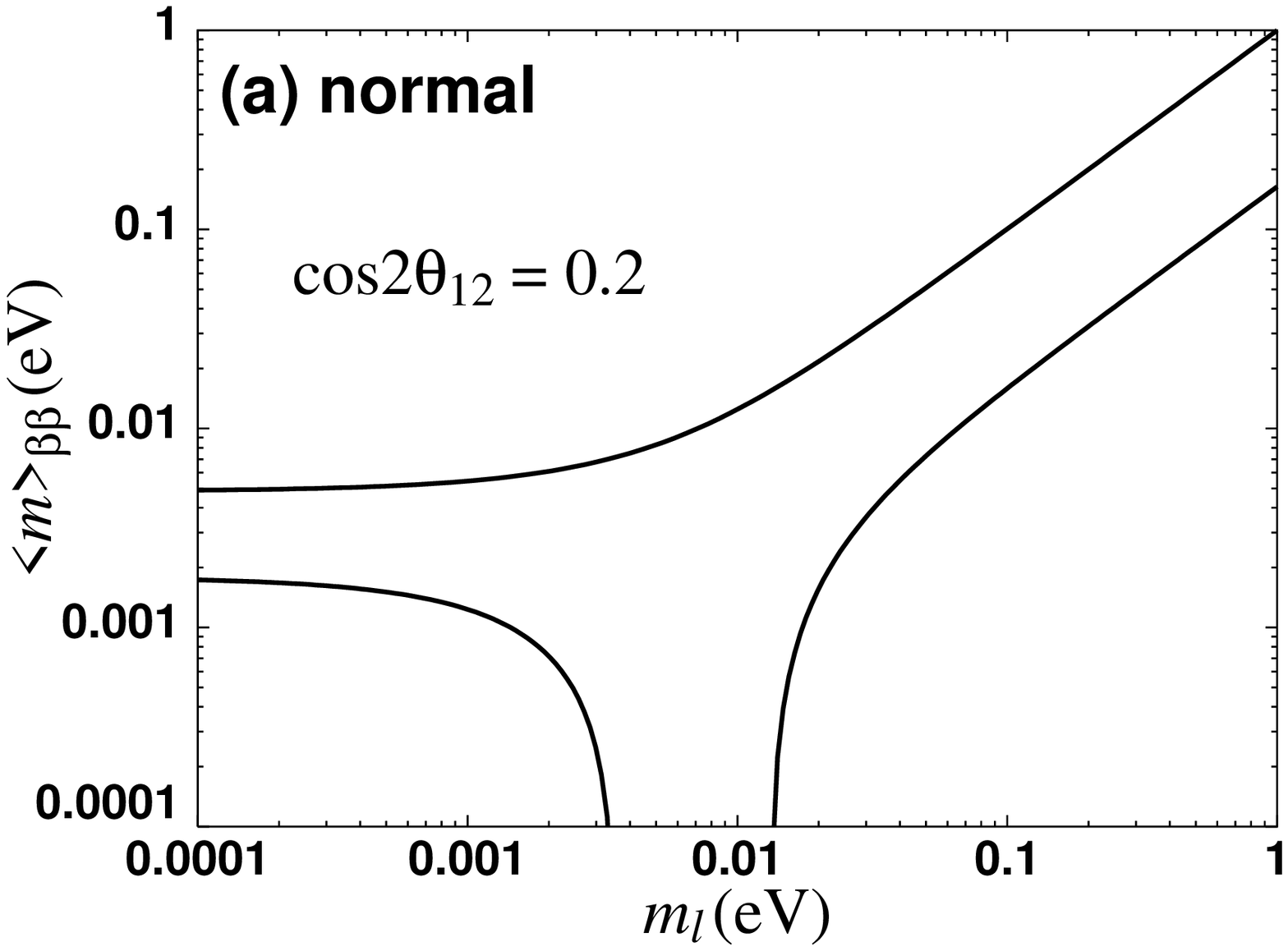} \ \ \
 \includegraphics[width=0.48\textwidth]{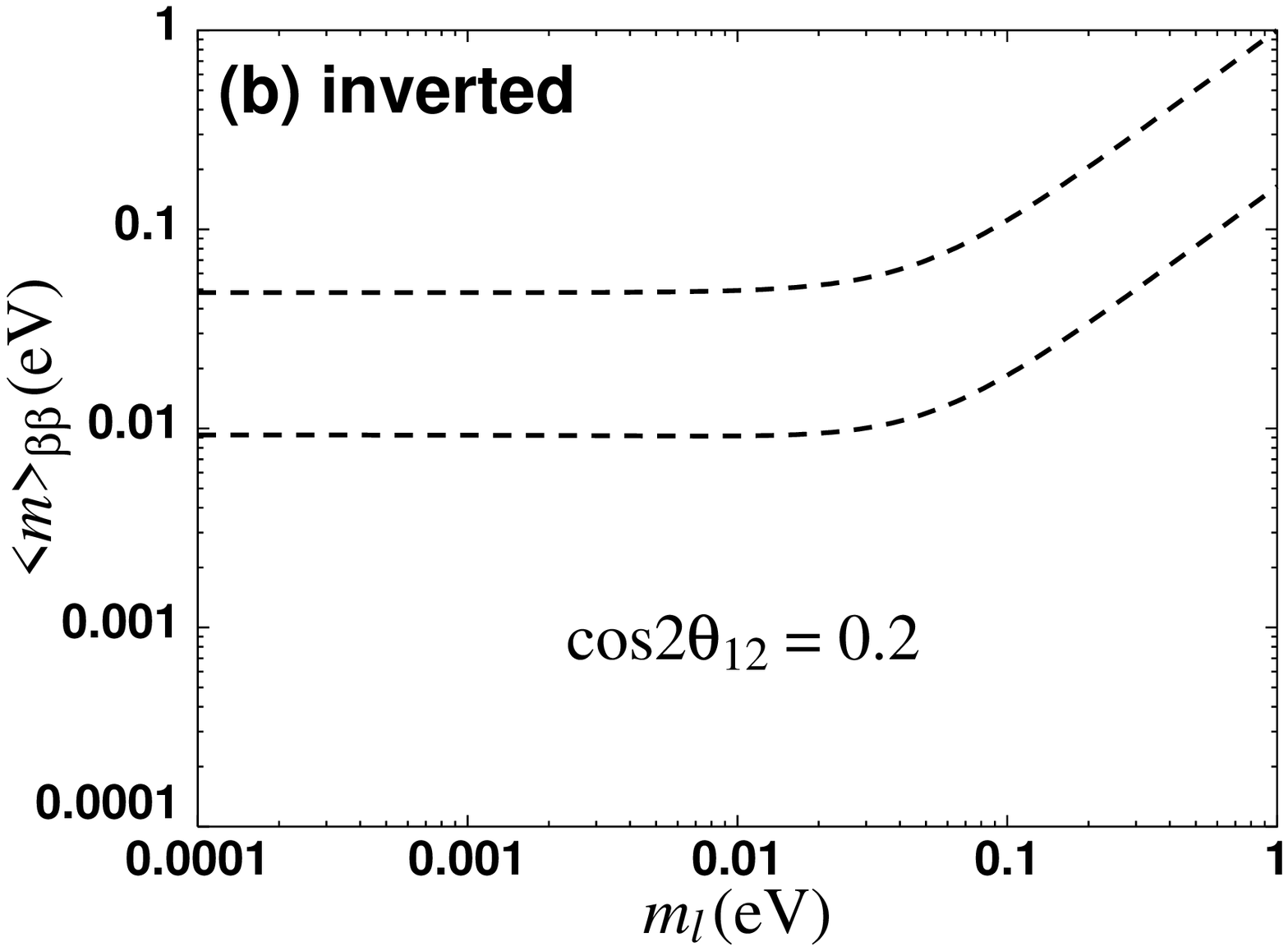}
 \caption{Shown are theoretical bounds on $\mbb$
with $\cos{2\theta_{12} = 0.2}$ which is the bound of LMA:
$0.2 < \cos{2\theta_{12}} < 0.5$.%
(See the text in Sec.~\ref{secttheo}.)
Figs.~(a) and (b) are for the normal and inverted hierarchy, respectively.
The region surrounded by those lines is allowed.
Note that $\mbb$ has the absolute lower bound,
$\simeq 0.05 \times \cos{2\theta_{12}}\,\eV$, for the inverted hierarchy
although does not for the normal one.}
 \label{theo}
\end{figure}

 By choosing phase factors and $\theta_{13}$
so as to maximize the right-hand side of (\ref{beta1}),
we obtain theoretical upper bounds on $\mbb$ as
\begin{eqnarray}
 \mbb
  \le \left(m_1 c_{12}^2 + m_2 s_{12}^2 \right) c_\CH^2
        + m_3 s_\CH^2
\label{beta2}
\end{eqnarray}
for the normal hierarchy, and
\begin{eqnarray}
 \mbb \le m_1 c_{12}^2 + m_2 s_{12}^2
\label{beta3}
\end{eqnarray}
for the inverted hierarchy,
where $s_\CH$ ($c_\CH$) is the largest (smallest) value
of $s_{13}$ ($c_{13}$) determined by the CHOOZ experiment.
 In the similar way a theoretical lower bound is obtained as
\begin{eqnarray}
 \mbb
  \ge c_\CH^2 \left| m_1 c_{12}^2 - m_2 s_{12}^2 \right|
  - m_3 s_\CH^2 .
 \label{beta4}
\end{eqnarray}
 Figs.~\ref{theo} present allowed regions on the \mbox{$m_l$-$\mbb$} plane
determined by (\ref{beta2}), (\ref{beta3}), and (\ref{beta4})
for fixed $\theta_{12}$.
 Fig.~\ref{theo}(a) shows that
$\mbb$ can be zero for the normal hierarchy.
 It occurs around
$m_l = s_{12}^2 \sqrt{\Delta m^2_{12}/\cos{2\theta_{12}}}$
which is obtained by setting the right-hand side of (\ref{beta4}) zero
with $\theta_\CH = 0$.
 In the inverted hierarchy,
almost maximal $\theta_{12}$ is necessary for $\mbb$ to vanish
because $m_1 \simeq m_2$,
but it is outside of the LMA region.

 On the other hand,
Fig.~\ref{theo}(b) shows that
$\mbb$ has the absolute lower bound for the inverted hierarchy
with the value of $\theta_{12}$.
 The approximate form of the absolute lower bound,
$\mbb \gtrsim 0.05 \times \cos{2\theta_{12}}\,\eV$, is extracted
by setting $m_l = \sqrt{\Delta m^2_{12}} = 0$
in the right-hand side of (\ref{beta4})
for the inverted hierarchy.
 Considering $0.2 \lesssim \cos{2\theta_{12}}$ of LMA,
$\mbb$ must be larger than about $0.01\,\eV$.
 Therefore, the inverted hierarchy is rejected
for Majorana neutrinos
if experiments show $\mbb \lesssim 0.01\,\eV$.

\section{Constraints on $m_l$ by a $\0nbb$ result\label{sectmass}}
\begin{figure}[t]
 \includegraphics[width=0.48\textwidth]{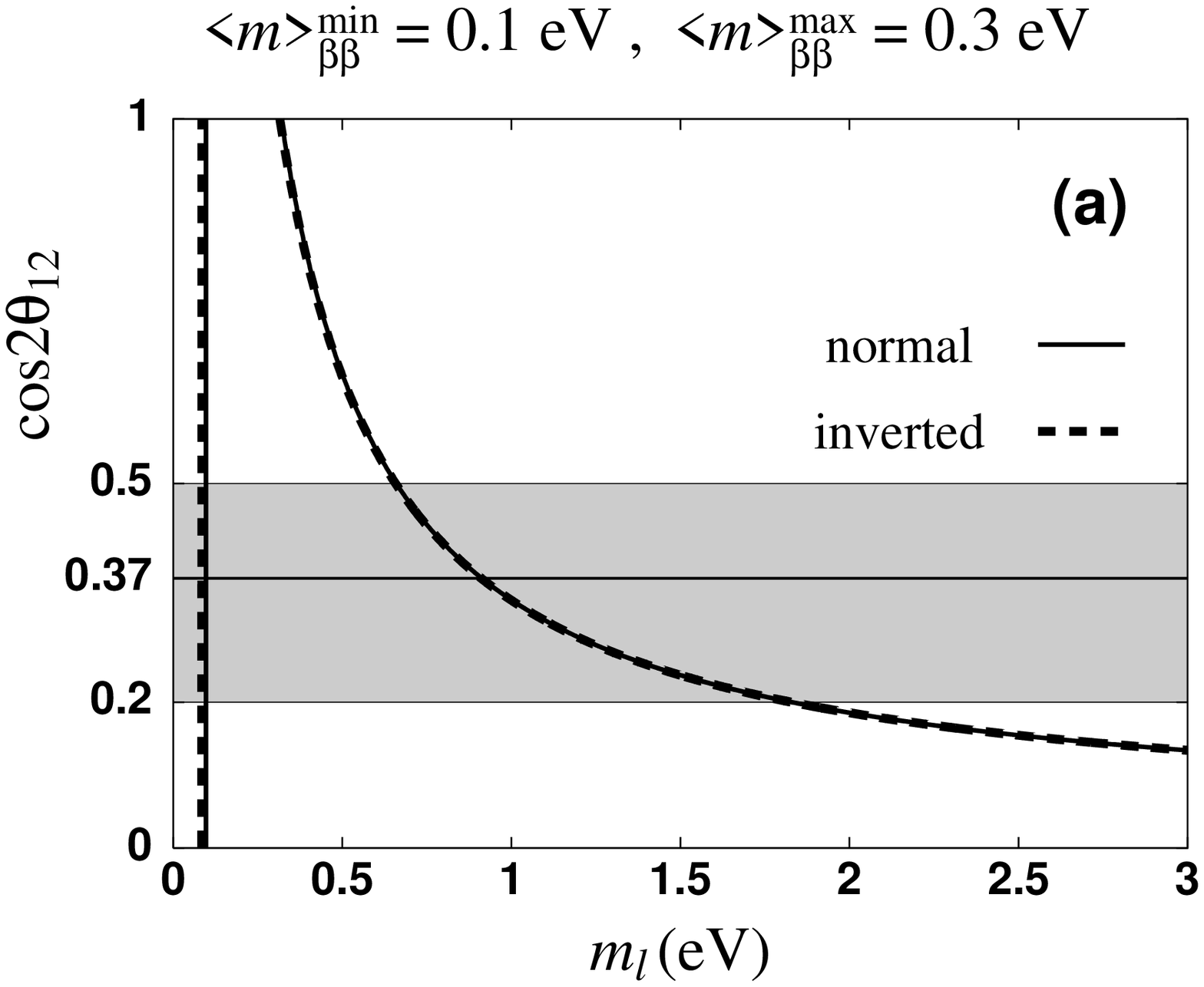} \ \ \
 \includegraphics[width=0.48\textwidth]{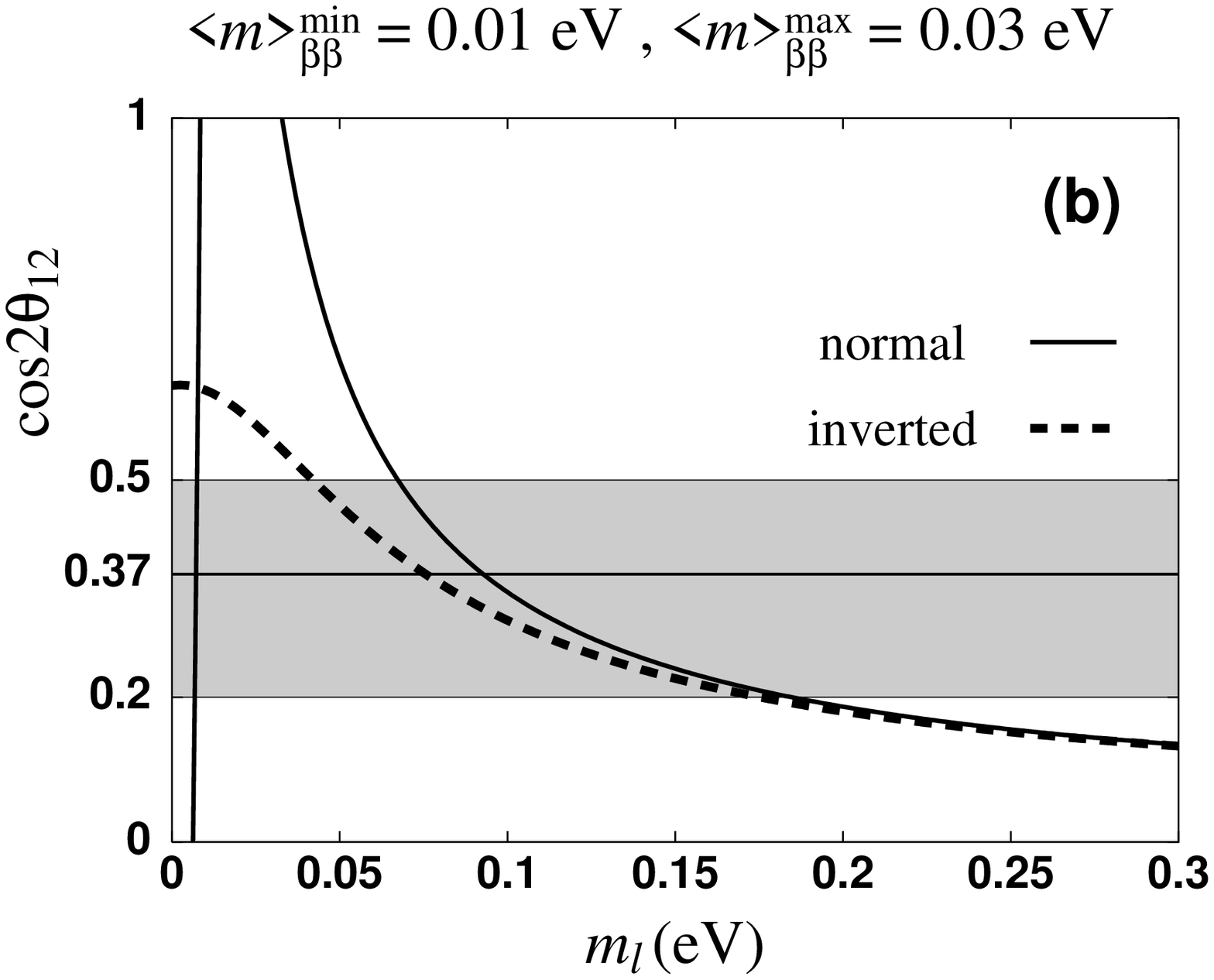}
 \caption{Presented are allowed regions
determined by two example cases of $\0nbb$ results.%
(See the text in Sec.~\ref{sectmass}.)
 The solid (dashed) lines are the bound
for the normal (inverted) hierarchy.
 LMA region is superimposed by the band of shadow.
 Note that large mixing is preferred by the inverted hierarchy.
 Note also that $m_l = 0$ can be allowed
even if $\mbb^\mi \neq 0$.}
 \label{massfig}
\end{figure}

 A neutrinoless double beta decay experiment puts
experimental bounds on $\mbb$ as $\mbb^\mi \le \mbb \le \mbb^\ma$
although $\mbb^\mi$ vanishes for a negative result.
 From now on, we try to utilize those experimental bounds.
 First, because the true value of $\mbb$ must be less than $\mbb^\ma$
and larger than the theoretical lower bound (\ref{beta4}),
we can construct an inequality
\begin{eqnarray}
 \mbb^\ma \ge \mbb
  \ge c_\CH^2 \left| m_1 c_{12}^2 - m_2 s_{12}^2 \right|
  - m_3 s_\CH^2 .
 \label{beta5}
\end{eqnarray}
 Similarly, other inequalities with $\mbb^\mi$ are obtained as
\begin{eqnarray}
 \mbb^\mi \le \mbb
  \le \left(m_1 c_{12}^2 + m_2 s_{12}^2 \right) c_\CH^2
        + m_3 s_\CH^2
\label{beta6}
\end{eqnarray}
for the normal hierarchy, and
\begin{eqnarray}
 \mbb^\mi \le m_1 c_{12}^2 + m_2 s_{12}^2
\label{beta7}
\end{eqnarray}
for the inverted hierarchy.
 These bounds determine allowed regions
on the \mbox{$m_l$-$\cos{2\theta_{12}}$} plane
which are shown in Fig.~\ref{massfig}.

 As an example $\mbb^\mi = 0.1\,\eV$ and $\mbb^\ma = 0.3\,\eV$
are considered in Fig.~\ref{massfig}(a).
 The bounds for the normal (solid lines)
and inverted (dashed lines) hierarchies are
almost same as each other
because the relevant scale of energy is large enough
compared with $\sqrt{|\Delta m^2_{23}|}$
so that the degenerate mass approximation applies.
 The bounds by (\ref{beta6}) and (\ref{beta7}) are almost vertical
because of small $\Delta m^2_{21}/m_l^2$.
 Thus, those bounds put a lower bound on $m_l$
almost independently of $\cos{2\theta_{12}}$,
which can be written approximately $\mbb^\mi \lesssim m_l$.
 The upper bound on $m_l$ can be extracted from (\ref{beta5}),
and the curve has an asymptotic line $|\cos{2\theta_{12}}| = t^2_\CH$.
 Consequently, the upper bound on $m_l$ does not exist
if $|\cos{2\theta_{12}}| < t^2_\CH$.
 Fortunately, it is not the case for LMA (gray band).
 Eventually,
if $\mbb^\mi$ and $\mbb^\ma$ are large enough
compared with $\sqrt{|\Delta m^2_{23}|}$,
the bounds on $m_l$ are roughly
 \begin{eqnarray}
  \mbb^\mi \lesssim m_l
   \lesssim \frac{1}{0.9\,\cos{2\theta_{12}}} \times \mbb^\ma .
   \label{mass}
 \end{eqnarray}
 The effect of nonzero $\theta_\CH$ is the reason
why the coefficient of $\cos{2\theta_{12}}$ is shifted from unity to 0.9.
 The most conservative upper bound
is obtained by $\cos{2\theta_{12}} = 0.2$
and then the coefficient of $\mbb^\ma$ is approximately 6.

 Next,
the case of $\mbb^\mi = 0.01\,\eV$ and $\mbb^\ma = 0.03\,\eV$
is considered in Fig.~\ref{massfig}(b) as another example.
 Future experiments are expected to be able to probe
such a small energy scale.
 The bounds in Fig.~\ref{massfig}(b) for two hierarchies
are no longer similar to each other
because $\Delta m^2_{23}/m_l^2$ is not negligible in the energy scale.
 The smaller $\mbb^\ma$ experiments achieve,
the larger the excluded region of small $\theta_{12}$
for the inverted hierarchy.
 It will be inconsistent with LMA if $\mbb^\ma \lesssim 0.01\,\eV$,
and then the inverted hierarchy is excluded
as discussed in the previous section.
 Another interesting feature of the bounds in Fig~\ref{massfig}(b)
is that $m_l = 0$ is allowed for the inverted hierarchy
although that is excluded for the normal hierarchy.
 It is because that $\mbb^\mi = 0.01\,\eV$ is less than
the theoretical upper bound on $\mbb$ at $m_l = 0$
for the inverted hierarchy. (See Fig.~\ref{theo}(b).)
 It can occur for the normal hierarchy also
if $\mbb^\mi$ is even smaller. (See Fig~\ref{theo}(a).)
 The sufficient conditions of $\mbb^\mi$ for the exclusion of $m_l = 0$
are obtained by using the theoretical upper bound on $\mbb$
with $m_l = 0$.
 Those are
 \begin{eqnarray}
  \mbb^\mi \gtrsim 0.005\,\eV \sim \sqrt{\Delta m^2_{12}}
 \end{eqnarray}
for the normal hierarchy, and
 \begin{eqnarray}
  \mbb^\mi \gtrsim \sqrt{|\Delta m^2_{23}|} \simeq 0.05\,\eV
 \end{eqnarray}
for the inverted hierarchy.

\section{A possibility of excluding Majorana neutrinos}
 Without information of the mass hierarchy,
the hypothesis of Majorana neutrinos can not be rejected
by $\0nbb$ experiments even if $\mbb = 0$. (See Fig.~\ref{theo}(a).)
 For the rejection, some help of other experiments is necessary.
 In this section
we consider a possibility of the rejection
with a help of ${}^3H$ beta decay experiments
which give direct measurement of the neutrino mass.
 The sensitivity limit is expected to be $0.3\,\eV$,
which is still in the degenerate mass regime,
in the future KATRIN experiment.
 Here, we consider the situation
that a positive result with $m_l > 0.3\,\eV$ is discovered.
 On the other hand, a negative result of $\0nbb$ experiments
put an upper bound $\mbb^\ma$ on $\mbb$,
and the bound is translated to an upper bound on $m_l$
as discussed in Sec.~\ref{sectmass}.
 Those two bounds on $m_l$ will be inconsistent with each other
if $\mbb^\ma$ becomes small enough.
 It means the rejection of the Majorana neutrino hypothesis.
 The critical value of $\mbb^\ma$ for the rejection
is obtained by setting the right-hand side of (\ref{mass})
equal to $0.3\,\eV$ which is the expected sensitivity
of the future ${}^3H$ beta decay experiment.
 Consequently, the necessary condition for the rejection
with $\cos{2\theta_{12}} = 0.2$ is
 \begin{eqnarray}
  \mbb^\ma < 0.05\,\eV .
 \end{eqnarray}
 The critical value $0.05\,\eV$ seems to be very important goal of
$\0nbb$ experiments
because it is accidentally the same as
$\sqrt{|\Delta m^2_{23}|}$ at the SK best fit value
where the difference between two hierarchies starts to arise.

\section{Placing a constraint on the Majorana phase\label{sectphase}}

\begin{figure}[t]
 \includegraphics[width=0.48\textwidth]{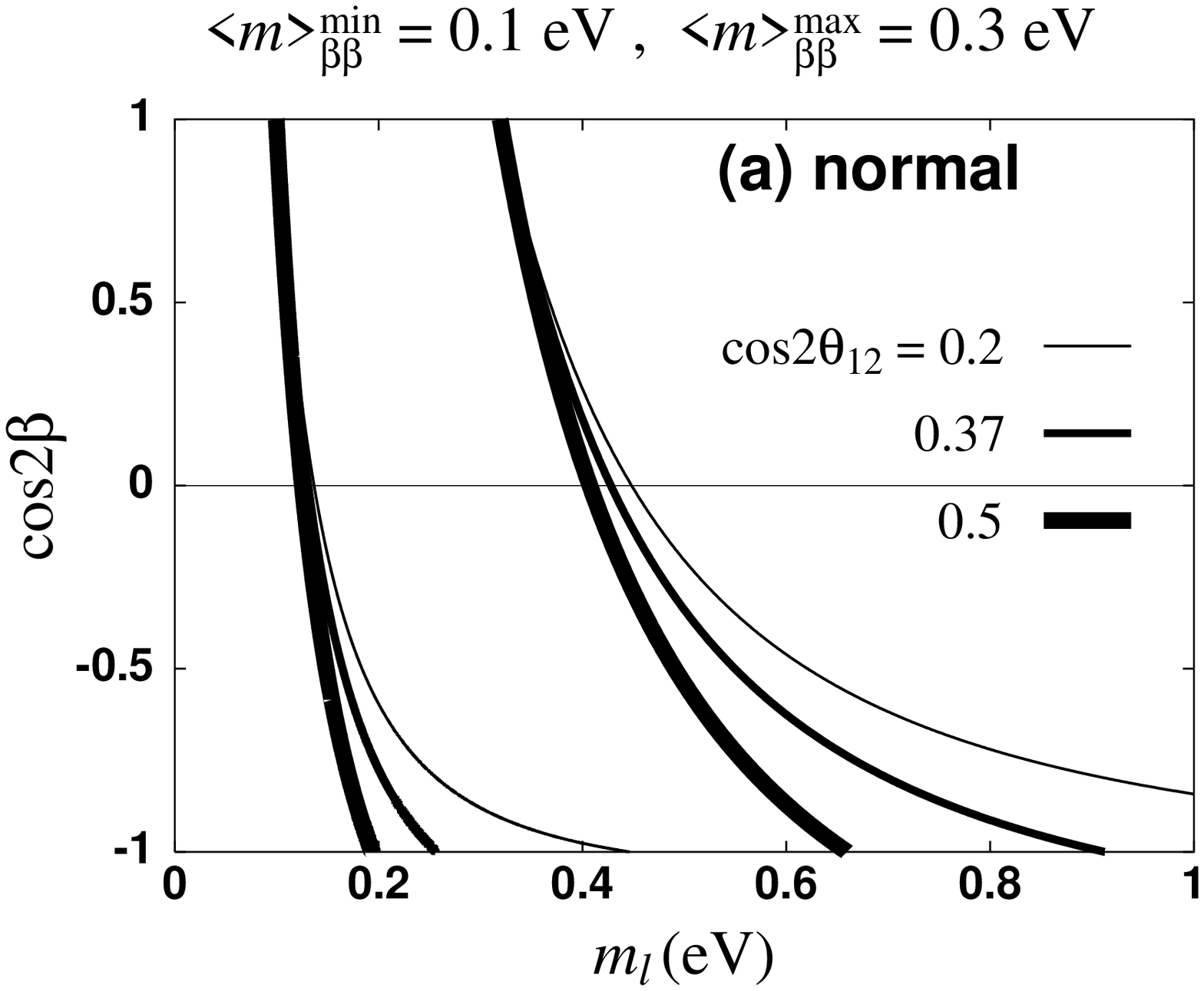} \ \ \
 \includegraphics[width=0.48\textwidth]{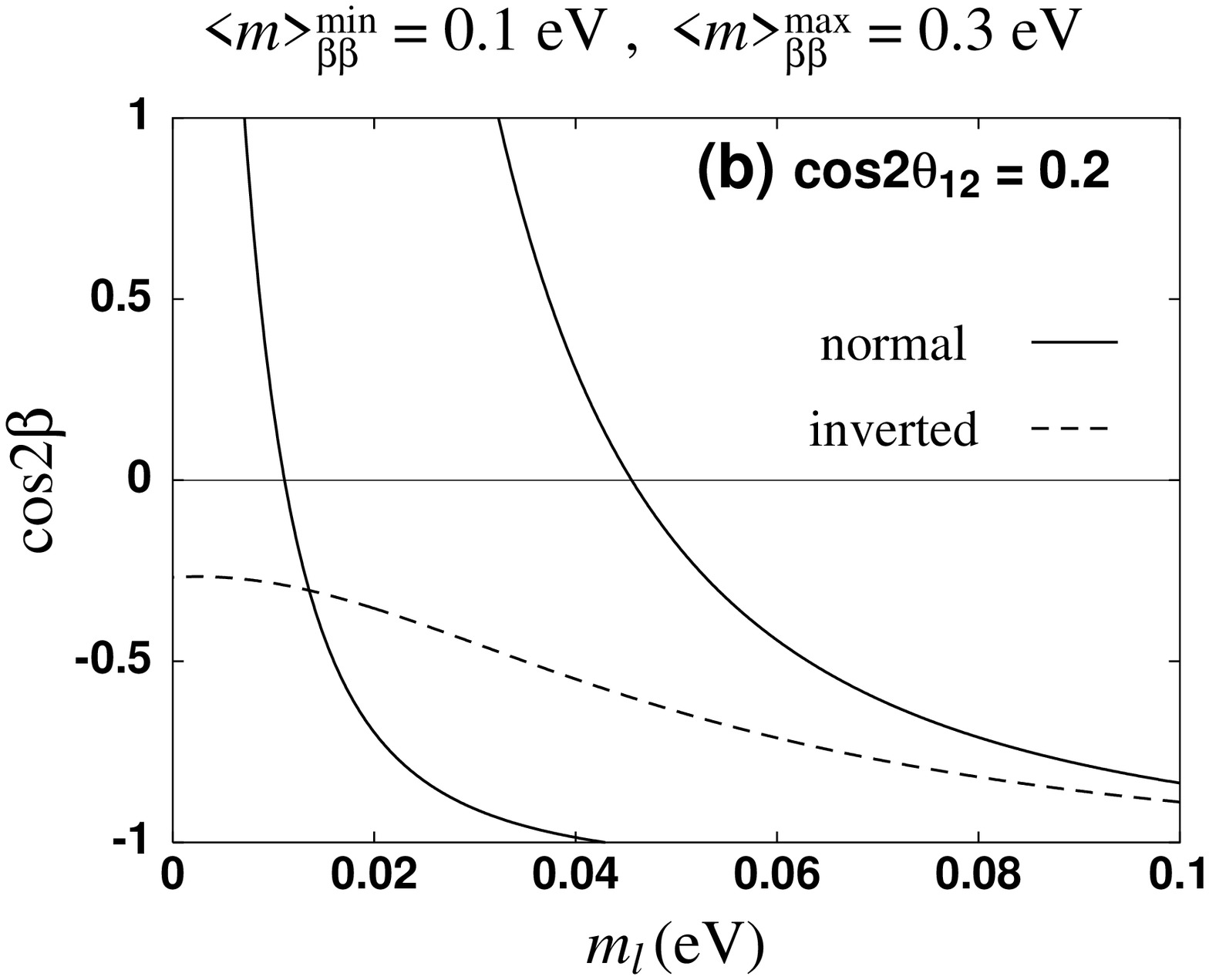}
 \caption{Shown are allowed regions
 by using two example cases of $\0nbb$ results.%
(See the text in Sec.~\ref{sectphase}.)
 Bold, normal, and thin lines correspond to
$\cos{2\theta_{12}} = 0.5$, 0.37, and 0.2
which are upper, best fit, and lower values of LMA, respectively.
 Solid (dashed) lines are for the normal (inverted) hierarchy.
 Insides of same line species are allowed.
 Presented bounds in Fig.~\ref{phasefig}(a) are
only for the normal hierarchy
because those bounds are almost independent of hierarchy
in the energy scale.
 For instance, if ${}^3H$ beta decay experiments show $m_l > 0.5\,\eV$,
we can put an upper bound on a Majorana phase
as $\cos{2\beta} \lesssim -0.25$ in Fig.~\ref{phasefig}(a).
 Lower bounds on $\cos{2\beta}$ seem to be difficult to obtain.}
 \label{phasefig}
\end{figure}

 When Majorana neutrinos are considered,
it is quite natural to hope that one day we will will be able to
access the Majorana phases.
 In general, it is impossible
even if we have precise values of $\mbb$, $m_i$, $\delta$, and mixing angles
because two unknown Majorana phases are involved in (\ref{beta1}).
 Nature, however, gives us a possibility to obtain
information of a Majorana phase $\beta$
at the price of the possibility for another phase $\gamma$.
 It can be seen in (\ref{beta1})
by noting that $\theta_{13}$ is tiny
and $\theta_{12}$ is not.

 By choosing $\theta_{13}$ and $\gamma-\delta$
appropriately in (\ref{beta1}),
we obtain an inequality which includes $\beta$ as
\begin{eqnarray}
  \mbb^\ma &\geq& \mbb\hspace*{85mm}\nonumber\\
  &\geq& c_\CH^2 \sqrt{m_1^2\,c_{12}^4 + m_2^2\,s_{12}^4
                      + 2\,m_1\,m_2\,c_{12}^2\,s_{12}^2\,\cos{2\beta}}
       - m_3\,s_\CH^2 .
\label{phase1}
\end{eqnarray}
 Similarly, we obtain
\begin{eqnarray}
 \mbb^\mi &\leq& \mbb\hspace*{80mm}\nonumber\\
 &\leq& c_\CH^2 \sqrt{m_1^2\,c_{12}^4 + m_2^2\,s_{12}^4
                      + 2\,m_1\,m_2\,c_{12}^2\,s_{12}^2\,\cos{2\beta}}
      + m_3\,s_\CH^2
\label{phase2}
\end{eqnarray}
for the normal hierarchy, and
\begin{eqnarray}
 \mbb^\mi \leq \mbb
 \leq \sqrt{m_1^2\,c_{12}^4 + m_2^2\,s_{12}^4
                      + 2\,m_1\,m_2\,c_{12}^2\,s_{12}^2\,\cos{2\beta}}
\label{phase3}
\end{eqnarray}
for the inverted hierarchy.
 Note that (\ref{beta5}), (\ref{beta6}), and (\ref{beta7})
are reconstructed by (\ref{phase1}), (\ref{phase2}), and (\ref{phase3})
with $\cos{2\beta} = 1$ or $-1$.
 Those inequalities determine allowed regions
on the $m_l$-$\cos{2\beta}$ plane for given $\theta_{12}$.

 We present the bounds on $\beta$ in Figs.~\ref{phasefig};
 The allowed regions are insides of the two lines with respective width.
 Presented in Fig.~\ref{phasefig}(a) are only for the normal hierarchy,
but bounds for the inverted hierarchy are almost the same as
those of the normal hierarchy
because the relevant energy scale is large enough
for the degenerate mass approximation to apply.
 Note that projecting the allowed region upon $m_l$ axis
results in the allowed region of $m_l$
which is nothing but the one obtained in Sec.~\ref{secttheo}.
 In Fig.~\ref{phasefig}(a),
the bounds determined by $\mbb^\ma$ exclude
large values of $\cos{2\beta}$ for large $m_l$,
and those bounds cross the line of $\cos{2\beta} = 1$
at around $m_l = \mbb^\ma$ almost independently of $\cos{2\theta_{12}}$.
 Thus, an upper bound on $\cos{2\beta}$ is extracted
if ${}^3H$ beta decay experiments show
that $m_l$ is larger than $\mbb^\ma$.

 On the other hand,
the bounds determined by $\mbb^\mi$ exclude
small values of $\cos{2\beta}$ for small $m_l$ in Fig.~\ref{phasefig}(a).
 In principle,
it is possible to put a lower bound on $\cos{2\beta}$
if other experiments give a stringent upper bound on $m_l$.
 In practice, however, it seems to be too difficult
because the values of $m_l$
that give a lower bound on $\cos{2\beta}$
tend to lie within too narrow region of too small energy scale;
 For example, the region is 0.1-0.2\,$\eV$ in Fig.~\ref{phasefig}(a)
for the most conservative case of $\cos{2\theta_{12}} = 0.5$.

 The difference between bounds for two hierarchies
can be seen in Fig.~\ref{phasefig}(b).
 For the inverted hierarchy,
the region of large $\cos{2\beta}$ is excluded for all $m_l$.
 When $\mbb^\ma$ becomes less than $0.01\,eV$,
whole region of $\cos{2\beta}$ is excluded for the inverted hierarchy.
 It means the exclusion of the inverted hierarchy
as discussed in Sec.~\ref{sectmass}.

\section{Summary}

 In this talk,
I presented an overview of constraints on neutrino masses and mixing
imposed by neutrinoless double beta decay ($\0nbb$).
 First, it was shown that $\mbb$ for the inverted hierarchy
has an absolute lower bound $\simeq 0.05 \times \cos{2\theta_{12}}\,\eV$,
and stringent experimental upper bounds on $\mbb$
can exclude the hierarchy.
 Second, constraints on the lightest neutrino mass $m_l$
was extracted by using a $\0nbb$ result.
 If the energy scales of a result is large enough
compared to $\sqrt{|\Delta m^2_{23}|}$,
the constraints are
$\mbb^\mi \lesssim m_l \lesssim \mbb^\ma / (0.9\,\cos{2\theta_{12}})$.
 Furthermore,
considering the expected sensitivity limit on $m_l$
of the future ${}^3H$ beta decay experiment
and assuming a positive result,
we uncovered a possibility of rejecting the Majorana neutrino hypothesis.
 It is necessary for the rejection
that $\0nbb$ experiments show $\mbb < 0.05\,\eV$.
 Next,
it was shown that
a bound on a Majorana phase $\beta$ can be placed
if a positive result of ${}^3H$ beta decay experiments
is obtained as $m_l > \mbb^\ma$.

 Finally, I would like to emphasize
the importance of precise determination
of the nuclear matrix elements $M^{0\nu}$.
 They are crucial to obtain stringent constraints
on neutrino properties in a manner fully utilizing
the accuracy of $\0nbb$ experiments.

\end{document}